\documentclass[prb,preprint]{revtex4}
\usepackage{graphicx}
\makeatletter
\usepackage{epsfig}
\def\@dotsep{4.5}

\begin{document}

\title{Calculations of the Energy Accommodation Coefficient for Gas-Surface Interactions }

\author{Guoqing Fan and  J. R. Manson}
 \affiliation{ Department of Physics and Astronomy, Clemson University, Clemson,
SC, 29634}
\date{\today}

\begin{abstract}

Calculations are carried out for the energy accommodation coefficient at a gas-surface interface using a recently developed classical mechanical theory of atom-surface collisions that includes both direct scattering and trapping-desorption processes in the physisorption well of the interaction potential.   Full three-dimensional calculations are compared with the available data for the accommodation of rare gases at a tungsten surface and good agreement is found for the heavier gases for which classical physics is expected to be valid at all measured temperatures.

\end{abstract}



\maketitle

\section{Introduction}  \label{intro}

The exchange of energy between a gas in contact with a surface is often characterized in terms of the  energy
accommodation coefficient.    Although the origins of the concept of an energy accommodation coefficient can be
traced to J. C. Maxwell~\cite{max} it is Knudsen who gave it a proper physical definition under what are now
known as the conditions of rarefied gas dynamics.~\cite{Knudsen-09,Knudsen-10,Knudsen-15,Knudsen-34}

The Knudsen energy accommodation coefficient has values that range from zero to unity, with a value of unity
arising if the gas achieves equilibrium with the surface after colliding with it and a value of zero implying
that no energy at all is transferred.

Early measurements of the accommodation coefficients for rare gases in contact with a tungsten surface were
carried out by Roberts, although it soon became apparent that his experiments were not carried out with
sufficiently clean surfaces and thus his data did not represent the values expected for the gas-surface
interaction with a clean metal.~\cite{roberts}  The work of Roberts did, however, stimulate early theoretical
investigations, especially for describing the interaction of He atoms with surfaces using quantum
mechanics.~\cite{Jackson,Devonshire-36,Devonshire-37} In the 1960s with the advent of high vacuum technology and
good surface cleaning techniques reliable measurements of the accommodation coefficient for rare gases on metal
surfaces became available from two different groups, that of Thomas et al.~\cite{Thomas1,Thomas2} and of
Kouptsidis and Menzel.~\cite{Menzel1,Menzel2} An extensive review of work pertaining to accommodation
coefficients and a very useful compendium of available experimental data has been presented by Saxena and
Joshi.~\cite{Saxena}  Other extensive reviews have been given by Goodman and
Wachman.~\cite{GoodmanProgSurfSci-1974,GoodmanErice,Goodman-76}

The purpose of this paper is to present calculations for the accommodation coefficients of the heavy rare  gases
with metal surfaces using a recently developed classical theory for atom-surface collisions that includes both
direct scattering and trapping-desorption processes.  Similar classical scattering theories have been applied
previously to calculations of the accommodation coefficient and reasonable agreement with measurements for the
heavy rare gases on clean tungsten surfaces was obtained, but these calculations included only direct scattering
processes and did not properly include trapping and subsequent desorption by the physisorption well of the
interaction potential.~\cite{Manson-Muis} The trapping-desorption fraction is that portion of an incident beam
of gas particles directed towards a surface that gets trapped by the physisorption well during the initial
collision process.  If this fraction remains in the physisorption well and does not go on to become permanently
adsorbed or chemisorbed (which is the expected case for rare gas atoms if the temperature is not too small)
these physisorbed atoms will eventually desorb and the standard assumption is that the trapping-desorption
fraction leaves the surface in a thermal energy distribution that is nearly in equilibrium at the temperature of
the surface.  Under such an assumption the trapping-desorption fraction is expected to enhance the accommodation
coefficient and cause it to have values closer to unity.  On the other hand, the direct scattering fraction
tends to exchange less energy with the surface and its contribution to the accommodation coefficient is expected
to cause it to have values less than unity.

The gas-surface scattering theory applied in this paper uses classical mechanics.  In the initial collision with
the surface a gas particle will either be scattered back into the continuum states (the direct scattering
fraction) or will be trapped in the physisorption well of the interaction potential.  Those particles that are
trapped can be subdivided into two classes, those that have negative total energy and those that have positive
total energy but are traveling at angles so close to the surface that they cannot escape from the well.  This
latter class is sometimes called the chattering fraction.  In the theory used here the trapped particles are
tracked as they make further collisions with the surface and with each subsequent collision a fraction remains
trapped but a fraction receives enough energy to escape into the continuum states.  These subsequent collisions
are treated with an iteration algorithm that can be carried out to very large numbers until virtually all of the
initially trapped particles have desorbed.  The theory has demonstrated clearly the conditions for the
trapping-desorption fraction to leave the surface in equilibrium, and it has also explained experimental data
for rare gas scattering under well-defined conditions for which both a direct scattering contribution and a
trapping-desorption fraction were observed as separate peaks in the energy-resolved scattering
spectra.~\cite{Fan-07}

As expected, for the accommodation of He and Ne at a tungsten surface, where quantum mechanics should be
dominant in the scattering process, the present classical theory is unable to explain the measured experimental
data.   However, good agreement with data is obtained for the heavy rare gases Ar, Kr and Xe.

\section {Theory}  \label{theory}

The energy accommodation coefficient $\alpha_E(T_S,T_G) $  is the ratio of the average energy exchanged  by a
gas in contact with a surface normalized to the maximum thermodynamically allowed energy that could be
exchanged.
\begin{equation}  \label{eqacc}
\alpha_E(T_S,T_G) \; = \; \frac{\overline{E_f}-<E_i>}{<E_f>-<E_i>} \; = \;
 \frac{\overline{E_f}-2 k_B T_G}{2 k_B T_S-2 k_B T_G}
 \; .
\end{equation}
In Eq.~(\ref{eqacc}) $T_G$ is the temperature of the gas, $T_S$ is the temperature of the surface,  and $k_B$ is
Boltzmann's constant,   $\overline{E_f}$ is the average energy of a gas particle after making a collision with
the surface.  The expression on the far right hand side of Eq.~(\ref{eqacc}) is obtained under the assumption
that both the gas and surface are in equilibrium at their respective temperatures, thus the average energy of
the incident gas is $<E_i> = 2 k_B T_G $  and the average energy of the gas if it should come into equilibrium
with the surface would be $<E_f> = 2 k_B T_S $.    These average energies are  obtained from the Knudsen
distribution  for a gas in equilibrium, sometimes called the flux-corrected Maxwell-Boltzmann distribution
\begin{equation}  \label{MB}
\frac{dP^{K}({\bf p}_i, T_G )}{ d E_i \: d \Omega_i}
\; = \;
\frac{E_i \cos \theta_i}{\pi (k_B T_G)^2}  \exp \left\{   \frac{-E_i}{k_B T_G}  \right\}
  \; .
\end{equation}

If the gas is initially in equilibrium then the average final energy after a collision with the surface is given
by
\begin{equation}  \label{ef}
 \overline{E_f} \; = \;
\int_0^\infty  d E_i \int_{2 \pi}  d \Omega_i \int_0^\infty  d E_f
\int_{2 \pi}  d \Omega_f  ~ E_f \;
\frac{dP^{K}({\bf p}_i, T_G )}{ d E_i \: d \Omega_i}
  \; \frac{dR({\bf p}_f,{\bf p}_i, T_S )}{ d E_f \: d \Omega_f}
\; ,
\end{equation}
where $ {dR({\bf p}_f,{\bf p}_i, T_S )} / { d E_f \: d \Omega_f} $  is the differential reflection coefficient
giving the probability per unit final energy and final solid angle that a gas particle initially in momentum
state ${\bf p}_i $ will make a transition to the state ${\bf p}_f $ as a result of the interaction with the
surface.  The differential reflection coefficient must obey the two conditions of unitarity and detailed
balancing, as does also the Knudsen distribution of Eq.~(\ref{MB}).  The condition of unitarity means that the
number of gas particles is conserved, i.e.,  for a given initial momentum state ${\bf p}_i $ the integral of the
differential reflection coefficient over all final energies and angles is normalized to unity.

It is convenient to define an accommodation coefficient that is a function of a
single temperature by taking the limit as the surface and gas temperatures approach the same value.  Thus
results in the equilibrium energy accommodation coefficient defined as
\begin{equation}  \label{EAC}
\alpha_E(T) \; = \;  \begin{array}{c} lim \\ T_G
\rightarrow T_S \rightarrow T \end{array} \;  \alpha_E(T_S,T_G)
 \; .
\end{equation}
All calculations in this paper will be for $ \alpha_E(T)$ since most experimental data for the accommodation of
rare gases on clean surfaces is reported in  this form.

Using the condition of detailed balancing, the temperature limit of Eq.~(\ref{EAC}) can be readily carried  out
leading to the final form
\begin{equation} \label{eeac}
\alpha_E(T)=\frac{1}{4(k_BT)^2}\int^\infty_0dE_i \int _{2 \pi}{ d
\Omega _i} \int^\infty_0dE_f \int _{2 \pi}{ d \Omega _f}  ~(E_f-E_i)^2 ~
\frac{dP^K({\bf p}_i; T_G)}{dE_i d\Omega _i} \frac{dR
({\bf p}_f ,{\bf p}_i; T_S )}{dE_f d\Omega
_f} \; .
\end{equation}

At this point the only remaining quantity needed for evaluating the accommodation coefficient is  the
differential reflection coefficient $ {dR({\bf p}_f,{\bf p}_i, T_S )} / { d E_f \: d \Omega_f} $. This provides
a complete description of the scattering process which means that it contains not only the direct scattering
arising from a single collision or a small number of collisions with the surface, but it also should contain the
contributions of those particles that are initially trapped and then subsequently desorbed.  The present authors
have recently developed a complete theory of  atom-surface scattering that includes both
contributions.~\cite{Fan-07}  This theory is based on an initial differential reflection coefficient for a
single surface collision $ {dR^0({\bf p}_f,{\bf p}_i, T_S )} / { d E_f \: d \Omega_f} $.   This initial
collision results in a  scattered intensity that consists of a fraction that is the direct scattering
contribution which leaves the surface and a fraction that is trapped in the physisorption well of the
interaction potential.  The trapped fraction is followed inside the well and when those particles have another
collision, some escape into the continuum and some remain trapped and go on to have further collisions with
similar consequences.  By dividing all trapped particles into a distribution of small energy and angular bins
they can be followed through many collisions until ultimately essentially all of them have escaped into the
continuum and desorbed.  This process can be written schematically as the following equation 
\begin{eqnarray} \label{iter}
\frac{dR({\bf p}_f,{\bf p}_i)}{d  {E}_f d \Omega_f}  ~=~ \frac{dR^0({\bf p}_f,{\bf p}_i)}{d  {E}_f d \Omega_f}
~+~ \int d E_b d \Omega_b ~ \frac{dR^0({\bf p}_f,{\bf p}_b)}{d  {E}_f d \Omega_f} ~ \frac{dR^0({\bf p}_b,{\bf
p}_i)}{d  {E}_b d \Omega_b}
\\ \nonumber
~+~ \int d E_b d \Omega_b ~ \frac{dR^0({\bf p}_f,{\bf p}_b)}{d  {E}_f d \Omega_f} ~ \frac{dR^1({\bf p}_b,{\bf
p}_i)}{d  {E}_b d \Omega_b} ~+~ \ldots
\\ \nonumber
~+~ \int d E_b d \Omega_b ~ \frac{dR^0({\bf p}_f,{\bf p}_b)}{d  {E}_f d \Omega_f} ~ \frac{dR^{n-1}({\bf
p}_b,{\bf p}_i)}{d  {E}_b d \Omega_b}
~,
\end {eqnarray}
where $ {dR^n({\bf p}_b,{\bf p}_i, T_S )} / { d E_b \: d \Omega_b} $ is the differential reflection coefficient
giving the distribution of particles remaining trapped in the bound states after $n$ collisions and the
intermediate integrations in the higher order terms are carried out only over angles and energies that pertain
to the trapped fraction.  The process described in Eq.~(\ref{iter}) is readily developed into an iterative
scheme in which the scattered distribution remaining in the well after the $n$th collision becomes the source
for the next collision.  The details of the calculation of the differential reflection coefficient of
Eq.~(\ref{iter}) are given in Ref.[\cite{Fan-07}] where it is shown that this procedure for calculating the
trapping-desorption fraction can not only explain the physical behavior of the trapping-desorption fraction, but
it can also explain experimental data for energy-resolved  scattering spectra of rare gases taken under
conditions in which there are distinct and clearly exhibited  features due to both direct scattering and
trapping-desorption.

The zeroth order differential reflection coefficient is chosen to be that due to a potential consisting of an
attractive square well in front of a smooth repulsive wall whose surface vibrates due to the thermal motion of
the substrate atoms.   The square well has depth $D$ and width $b$.  For a classical mechanical treatment a
square well is a reasonable approximation since it describes correctly the increase in energy and refraction
towards more normal incidence angles when a particle enters the physisorption potential. The width $b$ does not
affect the scattered intensities provided it is larger than the selvedge region of the surface, i.e., as long as
it is larger than the surface vibrational displacements.  However, trapping times are proportional to $b$. This
differential reflection coefficient has been shown to be useful in explaining a variety of atom-surface
scattering experiments and is given by~\cite{model1,model2,model3,model4,model5}:
\begin{eqnarray}  \label{Mx}
\frac{dR^0({\bf p}_f,{\bf p}_i; T_s)}{dE_f d\Omega _f}
\\ \nonumber
= \frac {m^2 {v_R^2 } \left| {\bf p}_f \right|} {4\pi ^3
\hbar ^2 p_{iz} {S_{u.c.} } N_D^0 } \left| {\tau _{fi} } \right|^2 \left(\frac{\pi }{k_B T_s \Delta E_0
}\right)^{3/2} \exp\left\{ { - \frac{(E_f - E_i  + \Delta E_0 )^2+{2v_R^2 {\bf P}^2} }{4k_B T_s \Delta E_0 } }
\right\} ~,
\end {eqnarray}
where $\Delta E_0  = ({\bf p}_f  - {\bf p}_i)^2 /2M $ is the recoil energy. $p_{iz}$ is the $z$ component of the
incident momentum,  $\left| {\tau _{fi} } \right|^2$ is the form factor of the scattering $N_D^0$ is the
normalization coefficient,  ${\bf P} = ({\bf P}_f  - {\bf P}_i)$ is the parallel momentum exchange, and
$S_{u.c.}$ is the area of a surface unit cell. The quantity $v_R$ is a weighted velocity of sound parallel to
the surface. It is expected to have a values of order of the Rayleigh phonon speed and can be calculated from
the complete surface phonon spectral densiy, however, it is usually taken to be a constant.~\cite{model1,model2,model3}

The amplitude $ {\tau _{fi} } $ of the form factor appearing in Eq.~(\ref{Mx}) is in general  the transition
matrix of the elastic interaction potential extended off of the energy shell to account  for inelastic scattering.  A good
approximation that has been extensively used is the first-order distorted wave Born approximation matrix element
which for a strongly repulsive surface is given by~\cite{Goodman-76}

\begin{equation}  \label{My}
\tau _{fi} ~=~ 4 p_{fz} p_{iz} / m
\end {equation}

The main numerical operations involved in carrying out calculations are the multiple integrals  involved in the
accommodation coefficient of Eq.~(\ref{eeac}) and in the iterative evaluation of the differential reflection
coefficient of Eq.~(\ref{iter}).  In each case these are six-dimensional integrations, although, if the surface
is azimuthally symmetric as is the case for the potential used here,  the accommodation coefficient reduces to a
five-dimensional integral.  The angular integrations are carried out using Gauss-Legendre algorithms and the
energy integrals with Gauss-Laguerre algorithms.

\section{Results}\label{result}

\begin{figure}
\includegraphics[width=6.0in]{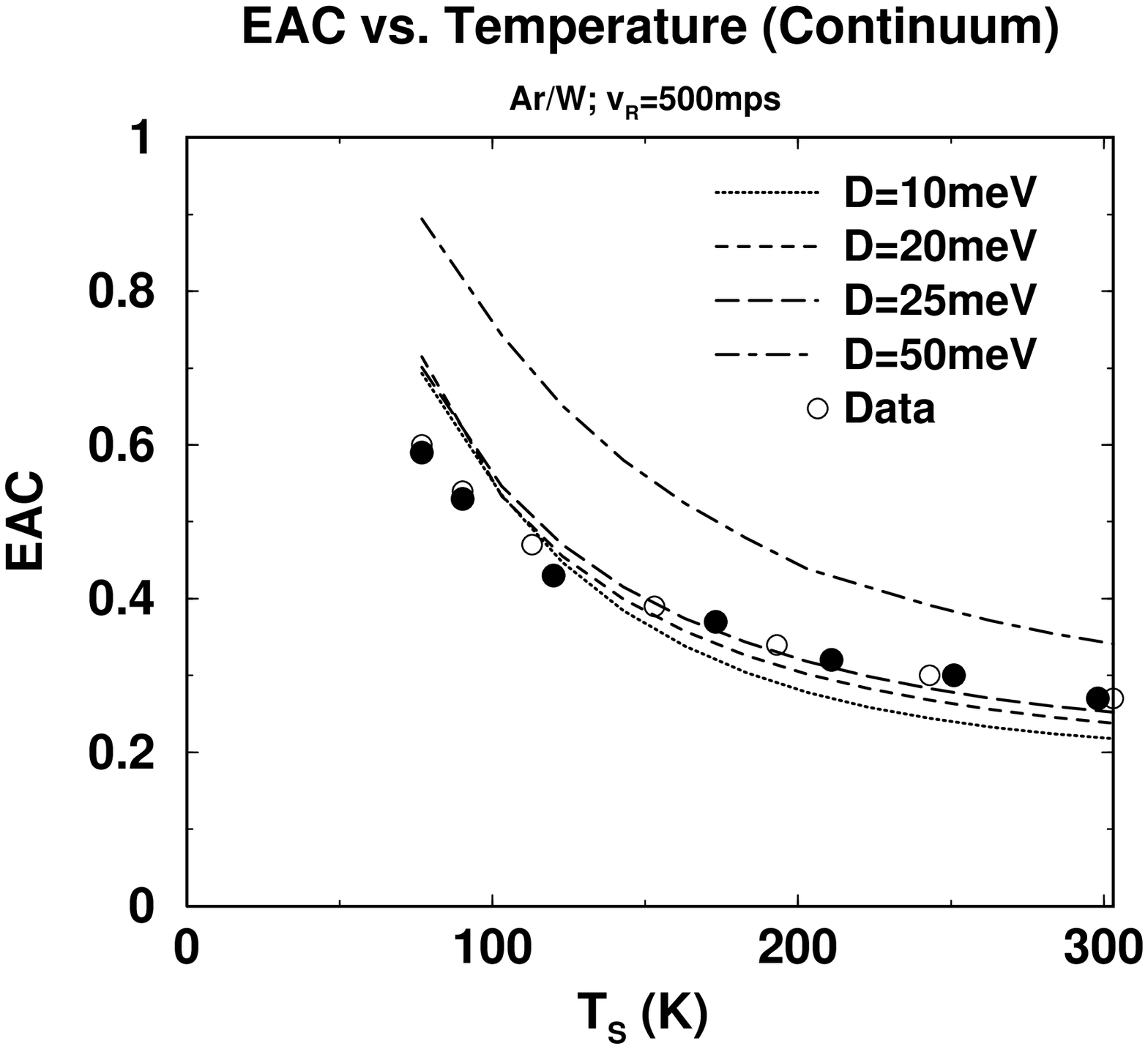}
\caption{The equilibrium energy accommodation coefficient $\alpha _E(T)$ as a function of the
temperature $T_S$ for Ar on a W surface.  Calculations four with different well depths as marked are
compared with the experimental data of Thomas {\em et al.}~\cite{Thomas1,Thomas2} (open circles) and Kouptsidis
and Menzel~\cite{Menzel1,Menzel2} (filled circles). }
 \label{arwv500}
\end{figure}

\begin{figure}
\includegraphics[width=6.0in]{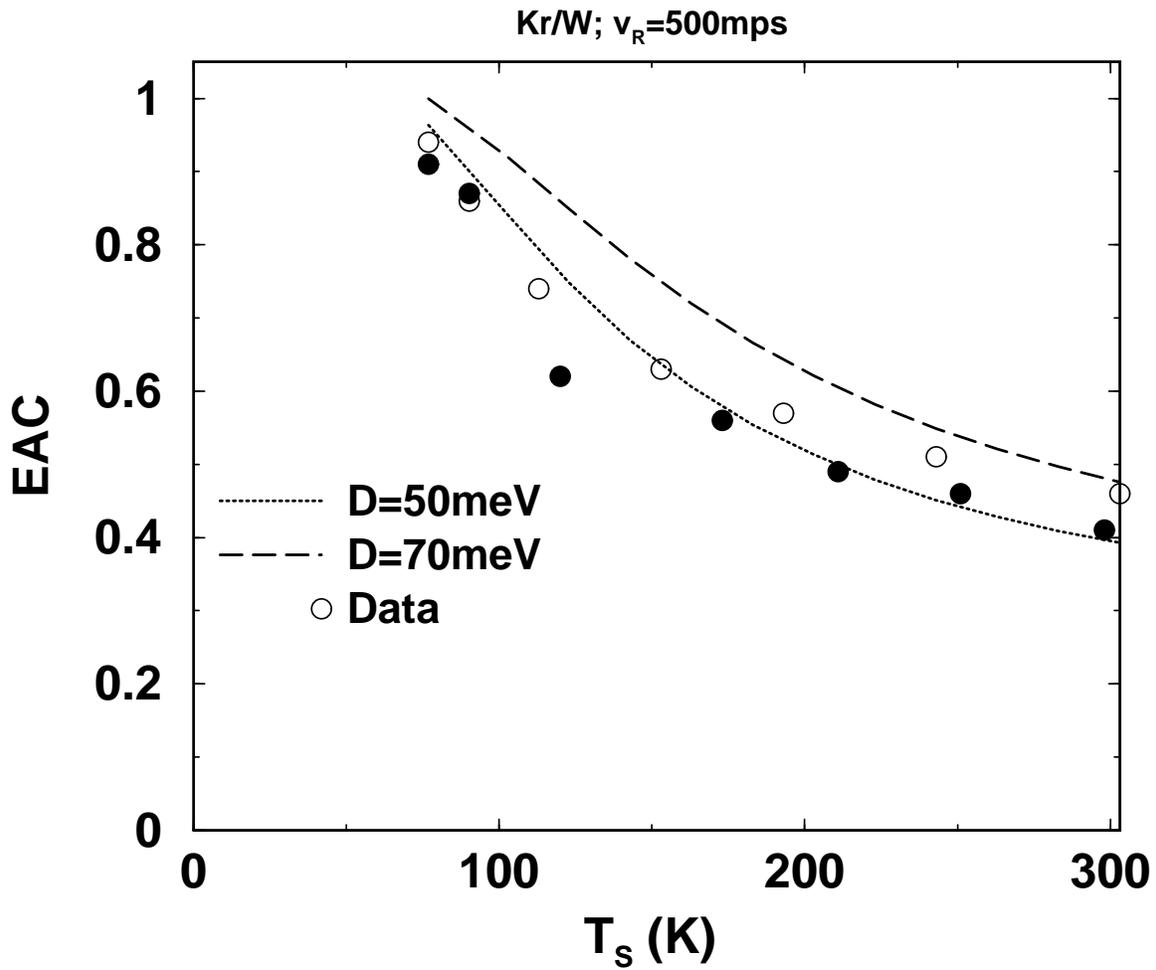}
\caption{ $\alpha _E(T)$ as a function of the
temperature $T_S$ for krypton gas in contact with a tungsten surface.   Two calculations with well depths of 50 and
70 meV are shown as marked and the data points are from the same sources and are denoted the same as in
Fig.~\protect \ref{arwv500}.}
\label{krwv500}
\end{figure}

\begin{figure}
\includegraphics[width=6.0in]{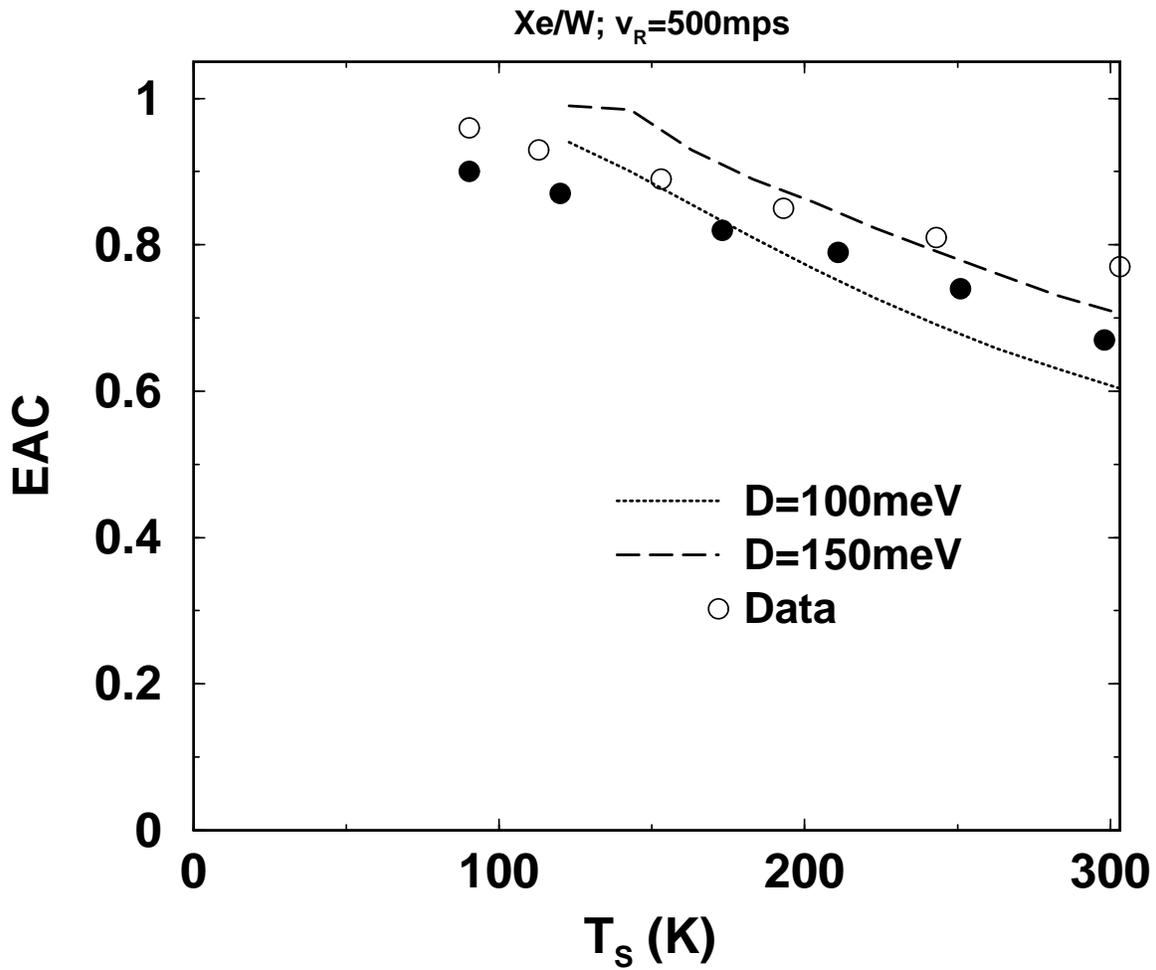}
\caption{The equilibrium energy accommodation coefficient $\alpha _E(T)$ as a function of the
temperature $T_S$ for xenon in contact with a tungsten surface.  Data are denoted as in Fig.~\protect\ref{arwv500}.
Calculations with well depths 100 and 150 meV are shown. }
\label{xewv500}
\end{figure}

Comparisons with experimental data for calculations using the theory and interaction potential  described above
are presented for the heavy rare gases in contact with a tungsten surface in
Figures~\ref{arwv500}-\ref{xewv500}.  The data from Thomas {\em et al.} are shown as open circles and the data
from Kouptsidis and Menzel are shown as filled circles.

Fig.~\ref{arwv500}  shows the measured equilibrium accommodation coefficient for Ar on W compared to  four
curves calculated with different well depths of 10, 20, 25 and 50 meV.  The velocity parameter is $v_R=500$ m/s.
The best agreement with the data is for a well depth of approximately $D=25$ meV.  A table of measured and
theoretically calculated well depths for the Ar/W system is given in Ref.~[\cite{Manson-Muis}] which shows that
this value of 25 meV is somewhat smaller than expected.  This table is based on values presented in
Ref.~[\cite{Ref68ofMuis-Manson}] and the measurements, primarily obtained from thermal desorption experiments,
range from 78 to 127 meV while calculated values are somewhat smaller ranging from 33 to 47 meV.  If the
velocity parameter $v_R$ is made somewhat larger the calculations for a larger well depth approach more closely
to the data at large temperatures, but the agreement at low temperatures becomes worse.  Although the well depth
predicted by these calculations is somewhat smaller than expected, it is considerably larger than the value of
15 meV used previously to fit the data with calculations based solely on the direct
scattering.~\cite{Manson-Muis}   Thus it becomes clear that including the trapping-desorption in the calculation
significantly increases the value of the accommodation.

Fig.~\ref{krwv500}  shows the accommodation coefficient data for Kr/W compared to calculations  carried out for
two different well depths, 50 and 70 meV. The best agreement with the data is for a well depth of approximately
$D=50$ meV.  Larger well depths lead to larger trapping-desorption fractions, and since the trapping-desorption
fraction is nearly in equilibrium this tends to enhance the accommodation coefficient.  As in the case for Ar/W,
calculations with  $v_R$ larger than 500 m/s will tend to decrease the accommodation coefficient for a given
well depth, but at the expense of poorer overall agreement with the data.    Estimated well depths for the Kr/W
system have been obtained only from thermal desorption experiments and these values range from 195 to 247 meV as
tabulated in Ref.~[\cite{Ref68ofMuis-Manson}].  Thus the value used here to give a best fit with the data is
small in comparison to the thermal desorption measurements, but again as for the Ar/W system it is twice as
large as the value obtained for calculations based only on direct scattering.~\cite{Manson-Muis}

Fig.~\ref{xewv500}  shows similar comparisons with data for the case of Xe/W. Calculations for two well depths,
100 and 150 meV, are shown and $v_R=500$ m/s. Both of these well depths are somewhat smaller than the
independent  measured thermal desorption value of 180 meV.~[\cite{Ref68ofMuis-Manson}].  It is interesting to
note that the calculations  for $D=150$ meV and  temperatures below $T_S=150$ K show that essentially all of the
gas atoms are trapped in the physisorption well and escape nearly in equilibrium which results in complete
accommodation, or $\alpha_E = 1$.

It is to be expected that a classical theory will not be adequate to describe the accommodation of the lighter
rare gases He and Ne on a surface of atoms as heavy as tungsten. Numerous treatments have shown that the
interaction of these gases with metal surfaces, especially for the case of He gas, is quantum mechanical in
nature and the scattering is dominated by elastic and single phonon inelastic processes.  A classical theory,
such as used here, cannot properly treat quantum mechanical processes and the present calculations predict
accommodation coefficient values that are much too large for He and Ne.  However, there are several quantum
theoretical treatments of the accommodation coefficient based on exchange of small numbers of phonons which
explain the measured values for the He/W and Ne/W system quite nicely.~\cite{Saxena,Goodman-76}

\section{CONCLUSIONS}\label{conclusion}

This paper presents calculations of the equilibrium accommodation coefficient for energy  exchange at a
gas-surface interface using a newly developed theory of atom-surface scattering in the classical limit that
treats both the direct scattering and the fraction of particles that are trapped and subsequently desorbed after
the initial collision with the surface.  This theory is applied to a relatively straightforward model of the
interaction potential, consisting of a strongly repulsive vibrating repulsive wall with an attractive square
physisorption well in front.  However, the theory treats the statistical mechanics of the scattering process
properly and is able to track all initially physisorbed particles until they eventually desorb.  This theory not
only describes correctly both the direct and trapping-desorption fractions, it has been used to explain measured
experimental data whose energy-resolved scattering spectra exhibit distinct features due to direct and
trapping-desorption events.  Thus, it is of interest to calculate the accommodation coefficient using this
theory to see if it explains the available data for energy transfer at a gas-surface interface.

A large amount of data exists for the accommodation of a variety of atomic and  molecular gases at
different types of surfaces.  However, the most carefully defined systems, both experimentally and theoretically,
are the rare gases accommodating at a tungsten surface.  Although data is available for all the rare gases
except radon, comparisons here are made only for the heavier rare gases.  This is because the light mass rare
gases, He and Ne, interact quantum mechanically and are not well explained by a purely classical theory.

This work can be viewed as a logical extension of an earlier paper by one of the authors in which  calculations
with a similar interaction potential model, but a theory that contained only the direct scattering component,
was applied to the energy accommodation coefficient.~\cite{Manson-Muis}   Thus, the present work when compared
to the previous results gives a clear indication of the contributions of the trapping-desorption fraction to the
accommodation.

Good overall agreement between calculations and measured accommodation coefficient data is obtained.   However,
the results do depend on the choice of the well depth and the velocity parameter that arises from the model of
the interaction potential.  Neither of these quantities has been well established for the interaction of heavy
rare gases with the tungsten surface.  The calculated values of the well depths that give the best agreement
with measurements tend to be somewhat smaller than estimates extracted from thermal desorption experiments,
although there are typically significant difference between such measurements in the cases where more than one
exists. In comparison with the previous calculations, however, the present work predicts well depths that
are significantly larger due to the influence of the trapping-desorption fraction.

The fact that this theory explains the available data for heavy rare gas accommodating at clean tungsten
surfaces, and the fact that state-to-state calculations explain recently available data for Ar scattering under
conditions where the energy-resolved spectra exhibited clear evidence for distinct direct scattering and
trapping-desorption features implies that it should be useful for predicting the energy accommodation for other
gas-surface systems.  In particular it should be able to predict the behavior of other systems as a function of
the experimentally accessible initial conditions such as temperature, well depth, gas particle mass and surface
mass.


\vspace{1cm}
\noindent
{\bf Acknowledgements} \\

\noindent
This work was supported by the US Department of Energy under grant number
DMR-FG02-98ER45704.

\newpage

\end{document}